\documentclass[aps,prl,twocolumn,showpacs,floatfix]{revtex4}
\usepackage{graphicx}
\usepackage{times}
\usepackage{nicefrac}
\usepackage{amsmath}
\usepackage{amsfonts}
\usepackage{amssymb}
\usepackage{amsthm}
\usepackage{epsf}
\usepackage{bm}
\usepackage{bbm}

\usepackage{dcolumn}
\newcolumntype{.}{D{x}{}{-1}}

\newcommand{\be}{\begin{eqnarray}}
\newcommand{\ee}{\end{eqnarray}}
\newcommand{\la}{\langle}
\newcommand{\ra}{\rangle}

\newcommand{\veps}{\varepsilon}

\newcommand{\bcalH}{{\mbox{\boldmath$\cal H$}}}

\newcommand{\balpha}{\bm{\alpha}}
\newcommand{\bnabla}{\bm{\nabla}}

\newcommand{\bgamma}{\bm{\gamma}}

\newcommand{\bfp}{{\bf p}}
\newcommand{\bfpr}{{\bf p^\prime}}
\newcommand{\bfq}{{\bf q}}

\newcommand{\aZ}{\alpha Z}

\begin{document}

\title{Many-electron QED corrections to the g factor of lithiumlike ions}
\author{A. V. Volotka,$^{1,2}$ D. A. Glazov,$^{2}$ V. M. Shabaev,$^{2}$
I. I. Tupitsyn,$^{2}$ and G. Plunien$^{1}$}

\affiliation{
$^1$ Institut f\"ur Theoretische Physik, Technische Universit\"at Dresden,
Mommsenstra{\ss}e 13, D-01062 Dresden, Germany \\
$^2$ Department of Physics, St. Petersburg State University,
Oulianovskaya 1, Petrodvorets, 198504 St. Petersburg, Russia \\
}

\begin{abstract}
A rigorous QED evaluation of the two-photon exchange corrections to the g factor
of lithiumlike ions is presented. The screened self-energy corrections are calculated
for the intermediate-$Z$ region and its accuracy for the high-$Z$ region is essentially
improved in comparison with that of previous calculations. As a result, the
theoretical accuracy of the g factor of lithiumlike ions is significantly increased.
The theoretical prediction obtained for the g factor of $^{28}$Si$^{11+}$
${\rm g}_{\rm th} = 2.000\,889\,892(8)$ is in an excellent agreement with the
corresponding experimental value ${\rm g}_{\rm exp} = 2.000\,889\,889\,9(21)$
[A. Wagner {\it et al.}, Phys. Rev. Lett. {\bf 110}, 033003 (2013)].
\end{abstract}

\pacs{31.30.J-, 31.30.js, 31.15.ac}

\maketitle
%
%
Highly charged ions provide not only a unique scenario for probing QED
effects in the strongest electromagnetic fields but also give access
to an accurate determination of fundamental physical constants and nuclear
parameters. In recent years, amazing progress has been made in the
experimental and theoretical investigations of the bound-electron g factor.
High-precision measurements of the ground state g factor of H-like
carbon \cite{haeffner:2000:5308} and oxygen \cite{verdu:2004:093002} and the
related theoretical calculations provided determination of the electron mass.
Recently, due to the substantial progress in the experimental accuracy of the
g factor of H-like carbon and silicon the mass of the electron is once again
substantially increased \cite{sturm:2014:467}. So far H- and Li-like silicon
ions represent the heaviest ions, where the g factor has been measured
\cite{sturm:2011:023002,sturm:2013:R030501,wagner:2013:033003}.
To date, these experiments provide the most stringent tests of the bound-state
QED (BS-QED) corrections in the presence of a magnetic field.
Accurate measurements of the g factor in few-electron ions, such as Li-like
calcium and B-like argon \cite{lindenfels:2013:023412}, are already anticipated.
The investigations of the few-electron ions unlike H-like ions provide also an
access to the many-electron QED corrections, which are represented
by a different facet of the QED diagrams.

The theoretical contributions to the g factor of Li-like ions can be separated
into one-electron and many-electron parts. The one-electron terms are
similar to the corresponding corrections to the g factor of H-like ions.
The many-electron contributions, which define the main difference between
the g factors of H- and Li-ions, were investigated in
Refs.~\cite{shabaev:2002:062104,yan:2002:1885,glazov:2004:062104,glazov:2006:330,
volotka:2009:033005,glazov:2010:062112}.
The many-electron contributions are mainly determined by the screened radiative
and the interelectronic-interaction corrections. For low-$Z$ ions, the screened
radiative corrections were obtained employing the perturbation theory to the
leading orders in $\aZ$ \cite{yan:2002:1885,glazov:2004:062104}. For
intermediate-$Z$ ions, the screening effect was evaluated by introducing the
effective screening potential in the QED calculations to all orders in $\aZ$
\cite{glazov:2006:330}. For high-$Z$ ions, the most accurate results for the
screened radiative corrections were obtained rigorously within a systematic
QED approach \cite{volotka:2009:033005,glazov:2010:062112}. The one-photon exchange
diagrams, which represent the interelectronic-interaction correction of the first
order in $1/Z$, were evaluated in the framework of QED in Ref.~\cite{shabaev:2002:062104}.
The second and higher-orders contributions of the interelectronic interaction were
calculated by means of the large scale configuration-interaction Dirac-Fock-Sturm
(CI-DFS) method in Ref.~\cite{glazov:2004:062104}. However, until now, for all values of $Z$
the theoretical uncertainty was determined by the interelectronic-interaction corrections
and for the intermediate-$Z$ region also by the screened self-energy corrections.
In the present Letter we report on the complete evaluation of the two-photon exchange
and the screened self-energy corrections in the framework of a rigorous QED approach
within an extended Furry picture.

%
In the extended Furry picture, to zeroth order we solve the Dirac equation with an
effective spherically symmetric potential treating the interaction with the external
Coulomb potential of the nucleus and the local screening potential exact to
all orders. This approach significantly accelerates the convergence of the
perturbation expansion. We use different types of the screening potential.
The simplest choice is the core-Hartree (CH) potential, which is created by
the charge density distribution of the two core electrons in the $1s$ state.
Other choices are the $x_\alpha$ potentials: Kohn-Sham (KS), Dirac-Hartree (DH),
and Dirac-Slater (DS), which were successfully employed in previous calculations
of highly charged ions \cite{sapirstein:2001:032506,sapirstein:2001:022502,
sapirstein:2002:042501,glazov:2006:330,volotka:2008:062507,kozhedub:2010:042513,
sapirstein:2011:012504}. Moreover, we have also employed the Perdew-Zunger (PZ)
potential \cite{perdew:1981:5048} and the local Dirac-Fock (LDF) potential
derived by inversion of the radial Dirac equation \cite{shabaev:2005:062105}.

\begin{figure*}
\includegraphics[width=0.8\textwidth]{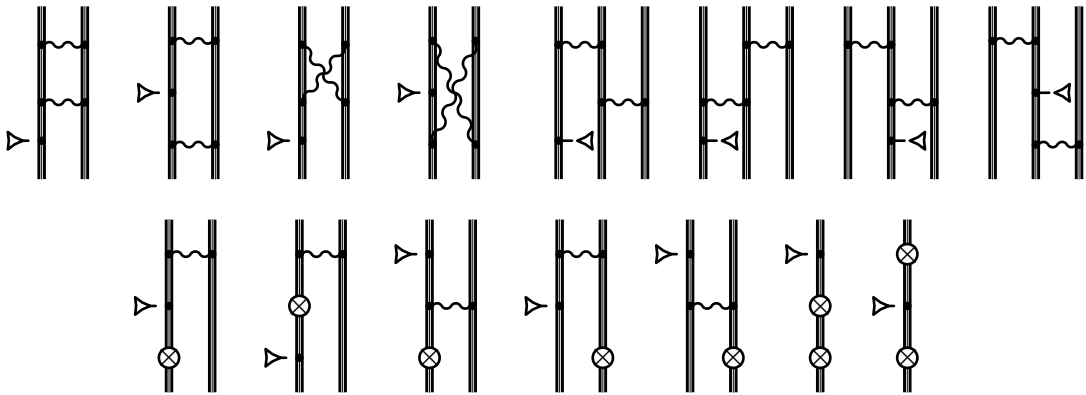}
\caption{Feynman diagrams representing the second-order interelectronic-interaction corrections
to the g factor in local effective potentials. The wavy line indicates the photon propagator and the
triple lines describe the electron propagators in the effective potential. The dashed line terminated
with the triangle denotes the interaction with the magnetic field. The counterterm diagrams are
depicted in the second line. The symbol $\otimes$ represents the extra interaction term associated
with the screening potential counterterm.}
\label{fig:inter}
\end{figure*}

Let us now turn to the evaluation of the two-photon exchange corrections to
the g factor of Li-like ions. These corrections are defined by diagrams of
third order in the QED perturbation theory. The corresponding diagrams are
presented in Fig.~\ref{fig:inter}.
The electron propagators in the figure have to be treated in the effective potential
(we indicate this diagrammatically via the triple electron line). In contrast to the
case of the original Furry picture, in the extended Furry picture the additional
counterterm diagrams appear. These diagrams are depicted in the second line in
Fig.~\ref{fig:inter}. They are associated with an extra interaction term represented
graphically by the symbol $\otimes$. Taking into account all possible permutations of
the one-electron states, in total, we have to evaluate 36 three-electron, 36 two-electron,
and 2 one-electron diagrams, respectively. All together these diagrams form the complete
gauge invariant set of the two-photon exchange contributions. The most difficult ones
are the 16 two-electron diagrams depicted in the first line in Fig.~\ref{fig:inter}.
Each of this diagram contains a three-fold summation over the complete Dirac spectrum
and an integration over the loop energy. Formal expressions for the diagrams in the
first line are similar to those derived for the corresponding calculation of the hyperfine
splitting and can be found in Ref.~\cite{volotka:2012:073001}. The formulas derived there
can be taken over but, instead of the hyperfine-interaction potential, we employ here
the interaction with a constant magnetic field and keep in mind that the Dirac spectrum
is now generated by solving the Dirac equation with the effective potential. The derivation
of the formal expressions for the diagrams of the second line is straight forward and
will be presented elsewhere. More details about the scheme of the numerical implementations
can be found in Ref.~\cite{volotka:2012:073001}. However, unlike the hyperfine splitting,
in the case of g factor the calculations are more involved due to the large cancellations
of various terms and poor convergence of the partial-wave expansion. Nevertheless, we have
substantially increased the accuracy of the all numerical integrations and extended the
partial-wave summation up to $\kappa_{\rm max} = 15$. For a consistency check, we performed
calculations both in Feynman and Coulomb gauges, and the results are found to be gauge
invariant with a very high accuracy.

In Table~\ref{tab:inter} the interelectronic-interaction corrections to the g factor of
Li-like silicon are given. The results are obtained with four different starting potentials:
Coulomb, core-Hartree, Perdew-Zunger, and local Dirac-Fock potentials. In the extended
Furry picture, the interelectronic interaction contributes already in the zeroth order,
due to the presence of the screening potential in the Dirac equation. The one-photon
(first order) and two-photon (second order) exchange corrections have been evaluated
to all orders in $\aZ$ in the framework of rigorous QED approach. The higher-order
corrections have been extracted from the calculations performed by means of the large-scale
configuration-interaction Dirac-Fock-Sturm method described in
Refs.~\cite{glazov:2004:062104,tupitsyn:2005:062503}. As it was expected, the employment
of the extended Furry picture increases the convergence of the perturbation expansion.
This allows us to reduce the absolute uncertainty of the higher order interelectronic-interaction
corrections.
\begin{table}
\caption{Interelectronic-interaction corrections to the ground-state g factor of
Li-like $^{28}$Si$^{11+}$ ion in various starting potentials in units $10^{-6}$.
\label{tab:inter}}
\begin{tabular}{lr@{}lr@{}lr@{}lr@{}l}                                            \hline\hline
& \multicolumn{2}{c}{Coulomb}
& \multicolumn{2}{c}{CH}
& \multicolumn{2}{c}{PZ}
& \multicolumn{2}{c}{LDF}
\\ \hline
Zeroth order  &       &        &   348.&267    &   321.&632    &   349.&636    \\
First order   &   321.&592     & $-$33.&549    &  $-$5.&990    & $-$33.&846    \\
Second order  &  $-$6.&876     &     0.&137    &  $-$0.&866    &  $-$0.&976    \\
Higher orders &     0.&085(22) &  $-$0.&046(6) &     0.&034(6) &  $-$0.&005(6) \\ [1mm]
Total         &   314.&801(22) &   314.&809(6) &   314.&810(6) &   314.&808(6) \\ \hline\hline
\end{tabular}
\end{table}
Finally, the rigorous evaluation of the two-photon exchange corrections and
the improved calculations of the higher-order terms allow us to significantly
increase the total accuracy of the interelectronic-interaction terms for all
ions under consideration. E.g., in the case of $^{28}$Si$^{11+}$ ion
the previous result was $0.000\,314\,903(74)$ \cite{glazov:2004:062104}, while
the present calculation yields $0.000\,314\,809(6)$, and in the case of
$^{208}$Pb$^{79+}$ ion instead of previous value $0.002\,140\,7(27)$
\cite{glazov:2004:062104} we now receive $0.002\,139\,34(4)$.

\begin{figure*}
\includegraphics[width=0.8\textwidth]{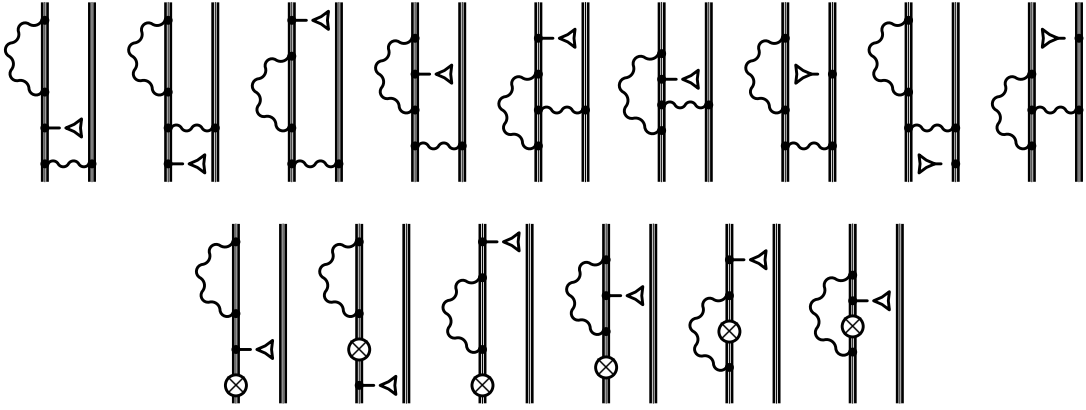}
\caption{Feynman diagrams representing the screened self-energy corrections to the
g factor in local effective potentials. The wavy line indicates the photon propagator and the
triple lines describe the electron propagators in the effective potential. The dashed line
terminated with the triangle denotes the interaction with the magnetic field. The counterterm
diagrams are depicted in the second line. The symbol $\otimes$ represents the extra interaction
term associated with the screening potential counterterm.}
\label{fig:scr-se}
\end{figure*}

Let us now turn to the screened self-energy corrections to the g factor of Li-like ions.
In Refs.~\cite{volotka:2009:033005,glazov:2010:062112} these corrections have been
rigorously evaluated only for the $^{208}{\rm Pb}^{79+}$ and $^{238}{\rm  U}^{89+}$ ions.
The reasons for this are twofold. First is the large numerical cancellations which occur
in the point-by-point difference. Second is the poor convergence of the partial-wave
expansion. In order to overcome these problems, we have performed the calculations in the
extended Furry picture and employed a special treatment of the many-potential terms.
The Feynman diagrams in the extended Furry picture corresponding to the screened self-energy
corrections to the g factor are presented in Fig.~\ref{fig:scr-se}.
The corresponding expressions derived in Refs.~\cite{volotka:2009:033005,glazov:2010:062112}
remain formally the same but keeping in mind that the Dirac spectrum is now generated by
solving the Dirac equation with the effective potential.  In the second line of
Fig.~\ref{fig:scr-se} the additional counterterm diagrams are depicted. The derivation of
the formal expressions for them is relatively simple and will be presented elsewhere.
The employment of the extended Furry picture allows us to substantially reduce the numerical
cancellations of different terms as well as to improve convergence of the partial-wave
expansion. However, in order to improve the convergence even further, we have employed
a specific treatment of some many-potential terms. The standard way to treat the vertex and
reducible corrections is to separate terms (zero-potential contributions) in which bound-electron
propagators are replaced by free propagators. The remaining many-potential terms being
ultraviolet finite are generally calculated directly in coordinate space \cite{blundell:1997:1857}.
However, for gaining better control over the partial-wave summation we separate also
the so-called one-potential contributions. In this way the one-potential terms are treated
in the momentum space. Such treatment of the one-potential term was applied in previous
calculations of the one-electron self-energy corrections to the g factor in
Refs.~\cite{persson:1997:R2499,beier:2000:032510,yerokhin:2002:143001,yerokhin:2004:052503,
yerokhin:2010:012502} and to the magnetic-dipole transition amplitude in
Ref.~\cite{volotka:2006:293}. Here, we extend this procedure to the evaluation
of the screened self-energy corrections to the g factor. Performing the analysis
of the convergence of the partial-wave expansion for different terms we have
found, that such treatment should be applied to the terms (C1) Eq.~(32), (H3) Eq.~(38),
(I1) Eq.~(47), and (I3) Eq.~(49) in Ref.~\cite{glazov:2010:062112}.
The corresponding one-potential contributions are given by the expressions:
\be
\Delta E_{\rm SQED}^{\rm SE(C1)(1)} = -8\pi i\alpha \sum_P (-1)^P
       \int\frac{d^3p\,d^3p'\,d^3q\,d^4k}{(2\pi)^{13}}\frac{1}{k^2} \nonumber\\
\times \bar{\psi}_a(\bfp)\gamma_\mu S_{\rm F}(p-k) \gamma_0 \Bigl[ V_{\rm eff}(\bfq)
       S_{\rm F}(p-k-q) \nonumber\\
\times T_0(\bfp-\bfpr-\bfq) + T_0(\bfq) S_{\rm F}(p-k-q) V_{\rm eff}(\bfp-\bfpr-\bfq)
       \Bigr] \nonumber\\
\times \gamma_0 S_{\rm F}(p'-k) \gamma^\mu \psi_{\zeta_{b|PaPb}}(\bfpr)
     + (a \leftrightarrow b) \,,
\ee
\be
\Delta E_{\rm SQED}^{\rm SE(H3)(1)} = -8\pi i\alpha
       \int\frac{d^3p\,d^3p'\,d^4k}{(2\pi)^{10}}\frac{1}{k^2} \bar{\psi}_a(\bfp) \nonumber\\
\times \frac{\partial}{\partial\veps_a}
       \Bigl[ \gamma_\mu S_{\rm F}(p-k) \gamma_0 V_{\rm eff}(\bfp-\bfpr)
            S_{\rm F}(p'-k) \gamma^\mu \Bigr] \nonumber\\
\times \psi_{\eta_a}(\bfpr)
     + (a \leftrightarrow b) \,,
\ee
\be
\Delta E_{\rm SQED}^{\rm SE(I1)(1)} = -4\pi i\alpha
       \int\frac{d^3p\,d^3p'\,d^4k}{(2\pi)^{10}}\frac{1}{k^2} \bar{\psi}_a(\bfp)\nonumber\\
\times \frac{\partial}{\partial\veps_a}
       \Bigl[ \gamma_\mu S_{\rm F}(p-k) \gamma_0 T_0(\bfp-\bfpr)
            S_{\rm F}(p'-k) \gamma^\mu \Bigr] \psi_a(\bfpr) \nonumber\\
\times \sum_P (-1)^P \la a b | I(\Delta) | Pa Pb \ra
     + (a \leftrightarrow b) \,,
\ee
\be
\Delta E_{\rm SQED}^{\rm SE(I3)(1)} = -4\pi i\alpha
       \int\frac{d^3p\,d^4k}{(2\pi)^{7}}\frac{1}{k^2} \nonumber\\
\times \bar{\psi}_a(\bfp) \frac{\partial^2}{\partial\veps_a^2}
       \Bigl[ \gamma_\mu S_{\rm F}(p-k) \gamma^\mu \Bigr] \psi_a(\bfpr)
       \la a | T_0 | a \ra \nonumber\\
\times \sum_P (-1)^P  \la a b | I(\Delta) | Pa Pb \ra
     + (a \leftrightarrow b) \,,
\ee
where $p = (\veps_a,\bfp)$, $p' = (\veps_a,\bfpr)$, $q = (\veps_a,\bfq)$,
$\Delta = \veps_a-\veps_{Pa}$, and the notation $(a \leftrightarrow b)$
stands for the contribution with interchanged labels $a$ and $b$;
$\gamma^\mu = (\gamma_0,\bgamma)$ are the Dirac matrices,
$S_{\rm F}(p) = (\gamma \cdot p - m)^{-1}$ is the free-electron propagator,
the interelectronic-interaction operator $I(\veps)$ and its derivatives are
defined in a similar way as in Ref.~\cite{glazov:2010:062112}, and $V_{\rm eff}$
is the effective potential being the sum of the nuclear and screening potentials.
$T_0$ is the operator of interaction with a constant magnetic field, which reads
in the momentum space:
\be
T_0(\bfp) = i \mu_0 (2\pi)^3 [\balpha\times\bnabla_{\bfp}\delta^3(\bfp)]\cdot \bcalH\,,
\ee
where $\mu_0 = |e|/2$ is the Bohr magneton and $\bcalH$ is the magnetic field
directed along the $z$ axis. The wave function $|\eta_a\ra$ is given by the expression:
\be
| \eta_a \ra = \sum_P (-1)^P
  \Biggl\{ | a \ra \Bigl[ \la \zeta_{b|PaPb} | T_0 | a \ra
                        + \la \zeta_{a|PbPa} | T_0 | b \ra \nonumber\\
                        + \la a b | I^\prime(\Delta) | Pa Pb \ra
                          \Bigl(\la a | T_0 | a \ra - \frac{1}{2} \la b | T_0 | b \ra \Bigr) \Bigr] \nonumber\\
         + | \xi_a \ra   \la a b | I(\Delta) | Pa Pb \ra
         + | \zeta_{b|PaPb} \ra   \la a | T_0 | a \ra \Biggr\}\,,
\ee
and the wave functions $|\xi\ra$ and $|\zeta\ra$ are defined similar as
in Ref.~\cite{volotka:2009:033005}.

The ultraviolet-finite one-potential terms given by Eqs.~(1)-(4) have been evaluated in the
momentum space. The corresponding expressions in the coordinate space have been subtracted
from the related many-potential terms by means of point-by-point difference. The partial-wave
expansion for the many-potential terms was terminated at $\kappa_{\rm max} = 15$, and the
remainder of the sum was estimated by a least-square polynomial fitting and by the
$\epsilon$-algorithm of the Pad\'e approximation. As a result, we have significantly
increased the accuracy of the screened self-energy correction. In the case of $^{28}$Si$^{11+}$
ion the previous result was $-0.000\,000\,218(46)$ \cite{glazov:2004:062104}, while the present
calculation yields $-0.000\,000\,242(5)$, and in the case of $^{208}$Pb$^{79+}$ ion instead
of previous value $-0.000\,003\,3(2)$ \cite{volotka:2009:033005,glazov:2010:062112} we
now receive $-0.000\,003\,44(2)$.

In Table~\ref{tab:total-g}, the individual contributions and the total values of the g factor
for Li-like silicon $^{28}$Si$^{11+}$, calcium $^{40}$Ca$^{17+}$, lead $^{208}$Pb$^{79+}$
and uranium $^{238}$U$^{89+}$ are presented together with the previously reported theoretical
results and the experimental value for the case of silicon.
\begin{table*}
\caption{Individual contributions to the ground-state g factor of Li-like ions and comparison
with the previously reported theoretical values as well as with the experimental result for the
$^{28}$Si$^{11+}$ ion.
\label{tab:total-g}}
\begin{tabular}{lr@{}lr@{}lr@{}lr@{}l} \hline\hline
& \multicolumn{2}{c}{$^{ 28}{\rm Si}^{11+}$}
& \multicolumn{2}{c}{$^{ 40}{\rm Ca}^{17+}$}
& \multicolumn{2}{c}{$^{208}{\rm Pb}^{79+}$}
& \multicolumn{2}{c}{$^{238}{\rm  U}^{89+}$}
\\ \hline
Dirac value (point nucleus) &   1.&998\,254\,751           &   1.&996\,426\,011       &   1.&932\,002\,904      &   1.&910\,722\,624      \\
Finite nuclear size         &   0.&000\,000\,003           &   0.&000\,000\,014       &   0.&000\,078\,57(14)   &   0.&000\,241\,62(36)   \\
QED, $\sim \alpha$          &   0.&002\,324\,044           &   0.&002\,325\,555(5)    &   0.&002\,411\,7(1)     &   0.&002\,446\,3(2)     \\
QED, $\sim \alpha^2$        &$-$0.&000\,003\,517(1)        &$-$0.&000\,003\,520(2)    &$-$0.&000\,003\,6(5)     &$-$0.&000\,003\,6(8)     \\
Interelectronic interaction &   0.&000\,314\,809(6)        &   0.&000\,454\,290(9)    &   0.&002\,139\,34(4)    &   0.&002\,500\,05(6)    \\
Screened self-energy        &$-$0.&000\,000\,242(5)        &$-$0.&000\,000\,387(7)    &$-$0.&000\,003\,44(2)    &$-$0.&000\,004\,73(3)    \\
Screened vacuum-polarization&   0.&000\,000\,006           &   0.&000\,000\,017       &   0.&000\,001\,53(3)    &   0.&000\,002\,55(5)    \\
Nuclear recoil              &   0.&000\,000\,039(1)        &   0.&000\,000\,061(2)    &   0.&000\,000\,25(35)   &   0.&000\,000\,28(69)   \\
Nuclear polarization        &     &                        &     &                    &$-$0.&000\,000\,04(2)    &$-$0.&000\,000\,27(14)   \\[1mm]
Total theory                &   2.&000\,889\,892(8)        &   1.&999\,202\,041(13)   &   1.&936\,627\,2(6)     &   1.&915\,904\,8(11)    \\
                            &   2.&000\,889\,909(51)$^a$   &   1.&999\,202\,24(17)$^b$&   1.&936\,628\,7(28)$^c$&   1.&915\,905\,7(41)$^c$\\
                            &   2.&000\,890\,005(87)$^b$   &     &                    &     &                   &     &                   \\
Experiment                  &   2.&000\,889\,889\,9(21)$^a$&     &                    &     &                   &     &                   \\
\hline\hline
\end{tabular}
\\
$^a$Wagner {\it et al.} \cite{wagner:2013:033003};
$^b$Glazov {\it et al.} \cite{glazov:2004:062104};
$^c$Glazov {\it et al.} \cite{glazov:2010:062112}.
\end{table*}
The screened self-energy and interelectronic-interaction corrections calculated in this Letter
allow to substantially increase the theoretical accuracy for all ions under consideration.
The other contributions to the g factor presented in Table~\ref{tab:total-g} were considered
in detail in our previous studies \cite{glazov:2004:062104,glazov:2006:330,volotka:2009:033005,
glazov:2010:062112}. Comparison with the experimental value for Li-like silicon ion provides
tests of relativistic interelectronic interaction on a level of $10^{-5}$, the one-electron
BS-QED on a level of 0.7\%, and the screened BS-QED on a level of 3\%. Thus, the current
studies provide the most accurate test of the many-electron QED effects in the case of g factor.
The further improvement of the g factor theory for Li-like ions requires at first the rigorous
evaluation of the three-photon exchange diagrams and the subsequent betterment of the screened
self-energy contribution for the intermediate-$Z$ region, and the one-electron two-loop and
nuclear recoil corrections for the high-$Z$ region.
%

The techniques and numerical methods developed can also be extended for the g factor
of B-like ions, where the corresponding studies can also lead to an independent
determination of the fine-structure constant \cite{shabaev:2006:253002}.
%

The work reported in this paper was supported by DFG (Grant No. VO 1707/1-2),
RFBR (Grants No. 13-02-00630 and 14-02-31316), and Saint Petersburg State
University (Grants No. 11.0.15.2010 and 11.38.269.2014).
D.A.G. acknowledges financial support by the FAIR -- Russia Research Center and
by the ``Dynasty'' foundation.
%

%
\end{document}